\documentclass[%
 reprint,
 amsmath,
 amssymb,
 aps,
 pra,
 showkeys,
]{revtex4-2}

\usepackage{graphicx}
\usepackage{dcolumn}
\usepackage{bm}
\usepackage{quantikz}
\usepackage[all]{xy} 
\usepackage{enumerate}
\usepackage{amsmath,amsthm,amssymb}
\usepackage{mathrsfs}
\usepackage[justification=justified]{caption}
\usepackage{xcolor}
\usepackage{verbatim}
\usepackage[normalem]{ulem}
\usepackage[rightcaption]{sidecap}
\usepackage{graphicx}
\usepackage{subcaption}
\usepackage{caption}
\usepackage{multirow}
\usepackage{epsfig}
\usepackage{pgfplots}
\pgfplotsset{compat=1.14}
\usepackage{float}
\usepackage{multirow}
\usepackage[colorlinks,citecolor=black,urlcolor=black,linkcolor =black,bookmarks=false,hypertexnames=true]{hyperref} 
\usepackage{relsize}
\usepackage{forest,adjustbox}
\usepackage{dsfont}
\usepackage{amscd}
\usepackage{boldline}
\usepackage{nccmath}

\theoremstyle{plain}

\newtheorem{assumption}[equation]{Assumption}
\newtheorem{lemma}[equation]{Lemma}
\newtheorem{proposition}[equation]{Proposition}

\newtheorem{claim}[equation]{Claim}
\newtheorem{definition}[equation]{Definition}
\newtheorem{remark}[equation]{Remark}

\newcommand{\Tr}{\operatorname{Tr}}

\DeclareMathOperator{\SU}{\mbox{SU}}

\DeclareMathOperator{\I}{\mathrm{I}}

\DeclareFontFamily{OT1}{pzc}{}
\DeclareFontShape{OT1}{pzc}{m}{it}{ <-> s*[1.2] pzcmi7t }{}
\DeclareMathAlphabet{\mathpzc}{OT1}{pzc}{m}{it}

\def\R{{{\mathbb R}}}

\newcommand\define[1]{\emph{\textbf{#1}}}

\begin{document}

\title{Geodesics of Quantum Feature Maps on the Space of Quantum Operators}

\author{Andrew Vlasic}
\affiliation{
Deloitte Consulting LLP
}

\date{\today}

\begin{abstract}
Recent advancements in the discipline of quantum algorithms have displayed the importance of the geometry of quantum operators. Given this thrust, this paper develops a rigorous geometric framework to analyze how the Riemannian structure of data, under the manifold hypothesis, influences the subspace of quantum gates induced by quantum feature maps. While numerous encoding schemes have been proposed in quantum machine learning, little attention has been given to how data geometry is deformed when mapped into the Lie group of special unitary operators. Addressing this gap, we assume a point cloud forms a smooth Riemannian manifold and formally construct the induced Riemannian geometry of a broad class of Hamiltonian quantum feature maps, which encompasses the majority of derived schemes. Starting from first principles, we derive analytic and, consequently, computational formulas for fundamental geometric measurements, including curvature, volume forms, and harmonic maps, providing tools for systematic deformation analysis. Notably, the derivations of the formulas elucidates how changes along paths in the data manifold interplay to changes in the associated subspace of special unitary operators, thereby indicating a direct geometric effect of data on quantum circuits. This framework establishes the mathematical validity required for principled analysis beyond heuristic ansatz and enables future research into geometry-aware quantum algorithm design.

\end{abstract}


\maketitle

\section{\label{sec:intro}Introduction}
\subsection{Overview}
This manuscript provides a precise and formal interplay between the curvature of the underlying manifold from the data and its effect on an encoding scheme. This formal connection also establishes a direct manner for computational analysis. To accomplish this task, it is necessary for a `ground-up' approach to establish a Riemannian geometry of the range of an encoding scheme by showing analogous characteristics and formulas found in literature on Riemannian manifolds on matrix Lie groups. Consequently, this work on foundation yields a computational and theoretical analytical framework for deeper and expanded research. 

Quantum machine learning (QML), motivated by the possibility that quantum systems can represent and process data in feature spaces that may be difficult to access classically \cite{havlicek2019supervised,cerezo2021variational}, has emerged as an active area of research both perspectives of near-term and full-tolerant computing. At the heart of many QML architectures is the quantum feature map: a data-dependent quantum operation that embeds classical information into a Hilbert space, or more generally, into the space of quantum states or gate operators \cite{schuld2019quantum,havlicek2019supervised}. This embedding is not merely a preprocessing step, for instance, in quantum kernel methods this steps dictates the induced feature states and hence the kernel leveraged by the learning model \cite{schuld2019quantum,havlicek2019supervised}. In variational QML architectures, the structure of the encoding interacts with the ansatz, cost function, and measurement scheme to shape expressivity and trainability \cite{cerezo2021variational,larocca2024review}. Thus, understanding quantum feature maps is an essential step for identifying when quantum models offer useful representations and for clarifying the conditions under which QML may provide a quantum advantage \cite{schuld2019quantum,havlicek2019supervised,cerezo2021variational}.

Encoding real-world data onto a quantum circuit, or a quantum feature map, is highly nontrivial, with many proposed schemes and methods to analyze the efficacy of the encoding, and of quantum circuits in general \cite{havlivcek2019supervised,schuld2019quantum,meyer2023exploiting,sim2019expressibility,schuld2021effect}. Through this vast literature, few papers fully mathematically explore the efficacy of an encoding technique and given data, especially from the perspective of the somewhat ambiguous term `information loss'. Considering the class of nontrivial Hamiltonian quantum feature maps, which encompass the vast majority of proposed encoding schemes, this manuscript derives the Riemannian manifold structure of the range, taken to be the Lie group of special unitary operators. Since the assumption that the range is a Lie subgroup does not in general hold, mathematical rigor is required to establish well-founded computational formulas and standard properties of Riemannian manifolds. In fact, this mathematical formality enables direct empirical analysis from closed formulas and further theoretical characterization, yielding deeper insight on the efficacy of given feature maps. 

Specifically, taking the manifold hypothesis of a given data point cloud, coupled with the assumption of a smooth Riemannian manifold, one may naturally ask how the data are deformed when encoded into the space of quantum operators. Note that the manifold hypothesis is typically taken in machine learning literature, and the smooth assumption can be obtained by `rounding-out' sharp edges. A Riemannian structure allows for a proper calculation of distances between points, the change in distances between points, the deformation of the local curvature, the change in volume, as well as a closed form for the generators of vector fields.

The consideration of the geometry of quantum circuits is quite natural as there already exists a geometric structure of quantum operators in quantum mechanics; see Bleeker \cite{bleecker2005gauge} and Helgason \cite{helgason1979differential}. In fact, this connection has been explored over the last few decades \cite{nielsen2005geometric,dowling2006geometry,brown2019complexity,ragone2024lie,wiersema2024here,goh2025lie}. Most notably, Wiersema et al. \cite{wiersema2024here} utilize the geometry of quantum operators to derive a more natural derivative for variational learning, and Ragone et al. \cite{ragone2024lie} describe the barren plateaus observed in variational learning to the dynamic Lie algebra generated from the quantum circuit. However, what is missing is the analysis of how the deformation of the geometry of a given point cloud---either from data collected in the natural sciences or general data science applications---effects a quantum circuit. This is contrary to classical modeling which has explored the importance of the data geometry \cite{gibilisco2010algebraic,friendly2013elliptical,marriott2024geometry,levenson2024advancing,vos2025geometry}. Therefore, we mitigate this shortcoming in the literature by deriving Riemannian techniques to analyze this interplay of the geometry of the data and the quantum operators within a quantum feature map.

Finally, addressing the necessity of mathematical rigor, for the range of encoding schemes as the space of quantum operators, the standard in the literature to derive the Riemannian geometry in this class of matrix Lie groups heavily leverages the group structure, connectedness, and a left-invariant metric to establish formulas and characteristics \cite{milnor1976curvatures,jost2008riemannian,hall2015lie}. Essentially, the tangent space at the identity matrix is first derived and the rest of the results follow. Since, in general, quantum feature maps do not hold a group structure, this negates many standard assumptions found in the literature. Therefore, it is necessary to work from a `ground-up approach', first establishing definitions before showing analogous results with respect to matrix Lie groups and necessary formulas for this manuscript.

\subsection{Outline}
Quantum features maps are typically derived via an ansatz. A notable exception is the paper by Meyer et al. \cite{meyer2023exploiting}, where the authors leverage the symmetric structure of the given data cloud. While intuitively important it is not very common with an arbitrary point cloud. Furthermore, when there exists multiple encoding options to represent the symmetry, the authors note one may just apply expressibility \cite{sim2019expressibility}. The note in the Meyer et al. \cite{meyer2023exploiting} brings us to well-used proposed techniques to analyze a quantum circuit, expressibility \cite{sim2019expressibility} and expressivity \cite{schuld2021effect}. While these method are sound, since the methods have not been fully mathematically vetted, the computational results are potentially ambiguous. With the goal of mathematical clarity, this manuscript posits that understanding how a quantum feature map deforms a manifold will yield deeper insight into the effects on a quantum circuit. However, a rigorous foundation is required to mathematically verify a Riemannian structure of the range before any effects from the data can be described.

The Riemannian manifold structure of quantum algorithms has been considered in the literature  \cite{dowling2006geometry,khan2018circuit,brown2019complexity,lewis2025geodesic}. The paths considered in the papers take the standard assumption that they generate a one-parameter subgroup when the identity is on the path. For example on of such a path, Dowling and Nielsen \cite{dowling2006geometry} consider a \textit{Hamiltonian representation} of a geodesic to derive a tangent vector to extract information about the geometry, but the geodesics are given from the global perspective. Lewis et al. \cite{lewis2025geodesic} leverage subgroup paths derive a learning algorithm with geodesics on $\SU(N)$, yielding an advantage over gradient descent by leveraging the natural geometric structure. A similar concept and algorithm is discussed in Wiersema et al. \cite{wiersema2024here}. We take a different approach and propose to leverage the structure of Riemannian manifolds to calculate how much a quantum feature map deforms a given point cloud, assumed to be an embedded manifold. 

Taking the assumption that a real data cloud is sampled from a smooth Riemannian manifold, call it $M$, we consider a quantum feature map of rotation gates $U:M \to \SU(2^N)$. We first derive the vector tangent space, the Riemannian metric on $U(M)$, and--- for completeness---derive the Lie algebra generated by the tangent space. From the metric, we are then able to calculate geodesics, curvature, and mathematically prove important properties. We then establish a closed analytic form to calculate volume and harmonic functions. Importantly, the derivations explicitly display how the curvature of the real manifold affects the curvature of the space of quantum gates.    

As previously noted, since the space $U(M)$ is not a Lie subgroup of $\SU(2^N)$---or a group---and so derivations and characterizations have to be constructed from the `base' level. Ergo, working first from definitions then mathematically establish properties assumed in the lemmas and propositions to build the mathematical structure for valid formulas for numerical or theoretical analysis.

The results given in the manuscript assumes a working knowledge of quantum feature maps, Lie groups and Lie algebras, as well as a little knowledge of Riemannian manifolds. However, the formulas rigorously derived are written in a manner  To assist the reader, references are given for definitions, equations, and important concepts. It is standard to use Einstein notation for summation within this discipline to simplify the notation. However, this notation can become cumbersome when details of the terms are needed, and as such, Einstein notation is not utilized. Furthermore, this manuscript does not address the data preparation step, which has the potential to deform the original manifold. However, there are many standard tools in Riemannian geometry one may leverage to answer any open questions.    

The structure of the manuscript is as follows. Section \ref{sec:lie-algebra-bundles} gives the closed form of the class of quantum feature maps analyzed, the closed form of the tangent space, as well as the Lie algebra generated by a tangent space, and, finally, properties of feature maps with matrices that commute or are noncommutative. The majority of the foundation is developed and provably displayed in Section \ref{sec:lie-algebra-bundles}. Section \ref{sec:geodesics} establishes closed form formulas for the metric, tensor curvature, sectional curvature, Ricci curvature, and ends with an illustrative example comparing two well-used feature maps and a restricted section of the Poincar\'e half-plane model. We finish the results with Section \ref{sec:cotangent-vol} where we give closed analytic forms to solve for the existence of a harmonic function. The crux of this derivation yields an illuminative interplay between the geometry of the data and the domain of a given quantum feature map. The manuscript ends with a discussion in Section \ref{sec:discussion}.

\section{\label{sec:lie-algebra-bundles}Tangent Spaces and Lie Algebras of Quantum Feature Maps}

\subsection{\label{subsec:feature-maps}Class of Quantum Feature Maps and the BCH Formula}

Quantum feature maps or encodings schemes, which we denote as $U:M\to \mathrm{SU}(2^N)$ for some $M \subset \mathbb{R}^n$, are quite often used in quantum machine learning (QML) literature as a linear combination of generative gates,
\begin{equation}
\label{eq:faeture-map}
U(p)=\exp\left(\sum_{k=1}^{n} f_{k}(p)L_{k}\right),
\end{equation}
where $L_{k}\in\mathfrak{su}(2^N)$, and $f_{k}:\R^{n} \to \mathbf{R}$ is smooth with $\mathbf{R} = [-\pi,\pi]$ or $\mathbf{R} = [0,2\pi]$. Observe that the range of $f_k$ is adjusted to encompass any values rotation and need not be surjective. For simplicity we define
\begin{equation}\label{eq:sum-of-operators}
L(p) := \sum_{k=1}^{n} f_{k}(p)L_{k}    
\end{equation}
so that Equation \eqref{eq:faeture-map} is simplified to $\displaystyle U(p) = \exp\left( L(p) \right)$. It will be convenient to reference $L(p)$, $L_k$, and $f_k$ throughout the manuscript.

To make Equation \eqref{eq:faeture-map} more tangible, from a well-known fact, multiple iterations of the Cartan decomposition of $\SU(2^N)$ yields that any element can be composed of tensor products of elements from $\SU(2)$ and $\SU(4)$; see Helgason \cite{helgason1979differential}. In other words, tensor products of Pauli operators and entanglement gates, like the CNOT gate. 

Feature maps in Equation \eqref{eq:faeture-map} are typically referred to as Hamiltonian encodings~\cite{SchuldPetruccione21}, or one-parameter subgroup homomorphisms~\cite{PBVP24} when the identity matrix is on a given path. For an explicit example of a quantum feature map, consider angle encoding~\cite{schuld21,schuld2019quantum, skolik2021layerwise}. Particularly, take 
$U_{\measuredangle}:[0,1]^{d}\to\mathrm{SU}(2^{d})$, where for $p = (p^1,\ldots, p^d)$ is given by 
\begin{equation*}
U_{\measuredangle}(p)=\bigotimes_{k=1}^{d} \exp\Big(  \arcsin( \sqrt{p^k} ) \cdot (-i\sigma_k)  \Big)
\end{equation*}
where $\sigma_{k}$ is a Pauli $\sigma$ gate acting on the $k^{\text{th}}$ qubit, ergo,
\begin{equation*}
\sigma_{k}=\mathbb{I} \otimes\cdots \otimes \mathbb{I} \otimes \sigma \otimes  \mathbb{I} \otimes \cdots \otimes \mathbb{I},
\end{equation*}
with $\mathbb{I}$ denoting the $2\times 2$ identity matrix. We remark that since the operators $\sigma_k$ commute the sum in Equation \eqref{eq:faeture-map} is equivalently the product of operators. 

However, not all unitary encodings in the context of QML are of the form~\eqref{eq:faeture-map}. For example, a feature map with creation and annihilation operators, motivated from fermions; see Montoya-Castillo and Markland \cite{montoya2018exact}. Also, there is the concept of data-uploading~\cite{PerezSalinas2020datareuploading,schuld2021effect,Jerbi2023} that use multiple layers, one layer being that of the Hadamard gate, which is a unitary operator and not a special unitary operator. For completeness, recall that unitary matrices can be written as a special unitary matrix multiplied by a global phase. Since the phase is global it has no effect on the measurement of the circuit. 

\begin{remark}
    The assumption that a feature map is of the form in Equation \eqref{eq:faeture-map}, at first glance, appears to be restrictive. However, the implementation of many encoding schemes, as well as applications in physics, use the first degree Suzuki-Trotter approximation in circuits, which agrees with Equation \eqref{eq:faeture-map} to the lowest order; see Nielsen and Chuang \cite{nielsen2001quantum} for Suzuki-Trotter. This also holds with implementations on various quantum modalities that are constrained to the respective native gates. Furthermore, while there are encoding schemes with layers of CNOTS, Hadamard gates, etc., the rotation values are constant, and hence, as well will see, have very little, if any, affect on the tangent space. Thus, this assumption encompasses the vast majority of encoding schemes.  
\end{remark}

Equation \eqref{eq:faeture-map} is given in such generality that it would be difficult to rigorously show properties. For example, if $f_{k}$ are constant for all $k$, or if $L_{k} \equiv 0$ for all skew-hermitian matrices then this trivial case will not hold many necessary properties. The assumption below negates any outlier encoding schemes. Observe that within this formal assumption, the domain of a feature map takes the manifold hypothesis of a given cloud of data, but one may take these results and adjust to manifolds that require charts for local coordinates.  

\begin{assumption}\label{ass:general-map}
For a quantum feature map given by Equation \eqref{eq:faeture-map}, we assume that the domain of a feature map is an embedded Riemannian manifold $(M,l)$ of dimension $m$, the map $p \mapsto U(p)$ is bijective, and $f_k$ is smooth for all $k$.  
\end{assumption}

For a multiple layer encoding, one is able to combine to an exponential term through multiple applications of the \define{Baker--Campbell--Hausdorff} (BCH) formula; see Hall \cite{hall2015lie}. For $X,Y \in \mathfrak{su}(N)$ and 
\begin{equation}\label{eq:commutator}
    [X,Y] = XY - YX
\end{equation}
as the \define{commutator} or Lie bracket, then the BCH formula has the form
\begin{equation}
e^{X}e^{Y}=e^{X+Y+\frac{1}{2}[X,Y]+\frac{1}{12}[X,[X,Y]]-\frac{1}{12}[Y,[X,Y]]+\cdots}.
\end{equation}
While the BCH formula will not be a major tool for the analysis, BCH serves as a basis for, and intuition to, the closed form of the derivative given in Equation \eqref{eq:derivative-exact}.

Now that we have established an explicit class of quantum feature maps with a computationally tangible encoding scheme, we move into deriving the tangent space of an arbitrary point in the range.

\subsection{\label{subsec:path-derv-bch}Paths and Derivatives}

Suppose that $\gamma:\R\to M$ is a smooth path in $M$, with $\gamma(0)=p$, and $U$ is a feature map of the form  Equation \eqref{eq:faeture-map}, then $U(\gamma(t))$ is a matrix-valued path. We will use this description to eventually show that all geodesic paths in $U(M)$ are exactly this form. Before we mathematically show this property in Proposition \ref{prop:geodesic-form}, we must formally establish a Riemannian structure. Although formal, the results are given in a computationally straight forward manner.   

To begin, we give a closed form for the derivative with respect to the input $t\in\R$. We utilize \cite[Theorem 5.4]{hall2015lie}, which states 
\begin{equation}
\label{eq:derivative}
\begin{split}
    \frac{d}{dt}e^{X(t)} &= e^{X(t)}\left( \frac{\I - e^{\mathrm{ad}_{X(t)}}}{\mathrm{ad}_{X(t)}} \right) \frac{dX}{dt} \\
    &:= e^{X(t)}\sum_{q=0}^{\infty}\frac{(-1)^{q}}{(q+1)!}\mathrm{ad}_{X(t)}^{q}\left(\frac{dX}{dt}\right)
    \end{split}
\end{equation}
for any smooth matrix-valued function $X(t)$. In the above notation, $\mathrm{ad}_X(Y) = [X,Y]$ denotes the \define{adjoint action}; the adjoint action is discussed in detail in  Section \ref{subsec:curvature}. Applying this formula to $U(\gamma(t))$, for $U$ in Assumption \ref{ass:general-map}, yields 
\begin{equation}\label{eq:derivative-exact}
\begin{split}
   & \frac{d U(\gamma(t))}{dt}\bigg|_{t=0}= \\& U(p)\sum_{q=0}^{\infty}\sum_{j=1}^{m}\sum_{k=1}^{n}\frac{\partial f_{k}}{\partial x^j}\bigg|_{p}\frac{d x^{j}}{dt}\bigg|_{t=0}\frac{(-1)^{q}}{(q+1)!}\mathrm{ad}^{q}_{L(p)}\left(L_{k}\right),
\end{split}
\end{equation}
where the $x^{j}$ are local coordinates on $M$ near the vicinity of the point $\gamma(0) = p$. For clarity, the exponential mapping of $U$ typically ensures the derivative is well-defined with some outlier cases generated from ambiguity, and the bijective assumption mitigates any ambiguity.   

Since this full description of the derivative is cumbersome---but perfect for computation---we define $\displaystyle \dot{ L } ( \gamma )(0) = \frac{d L( \gamma(t) )}{dt}\Big|_{t=0}$. From this simplified notation, we write the form for a tangent vector as
\begin{equation}\label{eq:derivative-feature-map}
    \begin{split}
        & \frac{d}{dt} \exp\Big( L( \gamma(t) ) \Big) \bigg|_{t=0} = e^{L( p )}\left( \frac{\I - e^{-\mathrm{ad}_{L( p )}}}{\mathrm{ad}_{L( p )}} \right) \frac{d L( \gamma )}{dt}\Big|_{t=0}\\
        & = U( p ) \left( \begin{split}
                    & \dot{ L } ( \gamma )(0) -\frac{ \big[L ( p ) , \dot{ L } ( \gamma )(0)  \big]}{2!} \\
       & + \frac{ \Big[L ( p ) , \big[L ( p ) , \dot{ L } ( \gamma ) (0) \big] \Big]}{3!} - \ldots
        \end{split}   \right) \\
        & := U( p ) \mathcal{L}(\gamma)(0).
    \end{split}
\end{equation}
Observe that the last equality defines the term $\mathcal{L}(\gamma)(0)$ as the repeated infinite sum from the BCH formula and will be utilized throughout the manuscript. 

Notice that if the collection of operators, $L_1, L_2, \ldots, L_{n}$, are pairwise commutative then Equation \eqref{eq:derivative-feature-map} simplifies to 
\begin{equation}\label{eq:derivative-feature-map-commutative}
    \frac{d}{dt} \exp\Big( L( \gamma(t) ) \Big) \bigg|_{t=0} = \exp\Big( L( p ) \Big) \dot{ L } ( \gamma )(0).
\end{equation}
Commutativity between all skew-hermitian matrices in the encoding scheme is an interesting property with, which we will see, important characteristics. Equation \eqref{eq:derivative-feature-map-commutative} induces Assumption \ref{ass:general-map} to include commutativity, which is described in the assumption below.

\begin{assumption}\label{ass:general-map-comm}
A quantum feature map given by Equation \eqref{eq:faeture-map} holds Assumption \ref{ass:general-map} and $L_{k}$ are all pairwise commutative.   
\end{assumption}

\subsection{Tangent Vector Space and Lie Algebras}\label{subsec:tanget-vector-lie-algebra}

From this closed form of a tangent vector and Assumption \ref{ass:general-map}, we are able to establish a tangent vector space. Before we discuss this vector space, observe that the operators applied in the Lie bracket in Equation \eqref{eq:derivative-exact} is a sum of skew-Hermitian matrices with the bracket recursively applied. Here, recall that for the roots of $\mathfrak{su}(2^N)$, say $\{ E_i \}$, for every $i \neq j$ there exists a $k$ such that $[E_i,E_j]=c^{k}_{ij}E_k$, for some $c^{k}_{ij} \in \mathbb{R}$. Then for the set of roots $\{E_{i'}\}$ that is the generator for the encoding scheme in Equation \eqref{eq:faeture-map}, if this set is not closed under the Lie bracket then the dimension of the tangent space would increase from repeated applications of the commutator. Thus, since the roots of skew-Hermitian Lie algebra are finite---that is when the Hilbert space is finite dimensional---we may condense the infinite sum of matrices into a finite sum of matrices, with each matrix a root in the Lie algebra $\mathfrak{su}(2^N)$. Of course, the coefficients of these matrices are convergent well-defined infinite sums.

Denote $\mathfrak{su}_{\mathcal{L}(p)}$ as the dynamic Lie algebra generated by the root operators in the sum $\mathcal{L}(\gamma)(0)$ for an arbitrary smooth path $\gamma$ with $\gamma(0)=p$. Combining the tangent vectors from all smooth paths $\gamma$, closing the set with scalars from the field $\mathbb{R}$ and closure under addition yields the equality 
\begin{equation}\label{eq:tangent-space-general}
        T_{U(p)} U(M) =  e^{L(p)}  \mathfrak{su}_{\mathcal{L}}. 
\end{equation}
To see why this equality holds, first note it is clear that $T_{U(p)} U(M) \subseteq  e^{L(p)}  \mathfrak{su}_{\mathcal{L}}$. For the other direction, a element in $e^{L(p)}  \mathfrak{su}_{\mathcal{L}}$ is $e^{L(p)}$ multiplied by a vector which is a linear combination of the roots generated from the derivative in Equation \eqref{eq:derivative-feature-map}. 

The claim is further explored and made more formal in the immediate Sections of \ref{subsec:commutative} and \ref{subsec:noncommutative-tangent}. This observation is not surprising since it follows a restricted structure of tangent spaces for a general operator $g\in \SU(N)$; see Hall \cite{hall2015lie}. 

Before proceeding to the next subsection, it would be prudent to discuss the form of a vector field. First, recall that a \define{vector field} is a smooth section from a manifold to the tangent bundle. Hence, $X:\Omega \to T\Omega$ with a smooth manifold $\Omega$. Of course, our vector field $X$ will be of the form $X:M \to TU(M)$.  However, since the focus of this manuscript is on data science applications, it is natural to consider the point cloud to be well-represented samples of an embedded bounded smooth manifold $M$. Needless to say, this is explicitly stated in Assumption \ref{ass:general-map}. Moreover, the smooth manifold assumption is also found in the natural sciences.

\subsection{\label{subsec:commutative} Commutative Feature Maps}

Considering the encoding scheme that satisfies Assumption \ref{ass:general-map-comm}, the commutative form of the chain rule, we show the natural isomorphic characteristic between tangent vector fields between Riemannian manifolds. However, as we will see, this property does not hold in the general case. Intuitively, there should exist an isomorphism between the tangent space of a given base point, $T_p M$ for $p\in M$, to the respective tangent space of $T_{U(p)}U(M)$. Specifically, for $\displaystyle \sum_{j=1}^m v^j\frac{\partial}{\partial x^j} \in T_p(M)$ we have $\displaystyle \exp\Big( L( p ) \Big) \sum_{i=1}^{n} \left( \sum_{j=1}^m v^j\frac{\partial f_i}{\partial x^j} (p) \right) L_i \in T_{U(p)}U(M)$. This logic leads to the following proposition, but first we need a lemma about a property of the Lie algebra generated by this vector tangent space.

\begin{lemma}\label{lemma:commutative-lie-algebra}
    Take a feature map where Assumption \ref{ass:general-map-comm} holds. Then the matrix basis elements of the tangent space are the roots that generate the function $L(\cdot)$. 
\end{lemma}
\begin{proof}
 Observe that $T_{U(p)}U(M)$ is a vector space generated from the roots identified within the BCH formula. We need only to show this vector space is closed under the commutator. For $\hat{U}_1, \hat{U}_2 \in T_{U(p)}U(M))$, since the skew-Hermitian matrices in the encoding scheme commute, $[\hat{U}_1, \hat{U}_2] = 0$. Thus $[\hat{U}_1, \hat{U}_2] \in T_M(U(p))$.
\end{proof}

While reading through the proof in the proposition below, note hat the proof of Lemma \ref{lemma:commutative-lie-algebra} will be doing a lot of heavy lifting.

\begin{proposition}\label{prop:commutative-tangent-space}
    Let $M$ be a Riemannian manifold and a quantum feature map of the form in Equation \eqref{eq:faeture-map} where Assumption \ref{ass:general-map-comm} holds. Then for all $p\in M$,
    \begin{equation}\label{eq:commutativity-iso}
        T_{p}M \cong T_{U(p)} U(M)
    \end{equation}
\end{proposition}
\begin{proof}
    Take an arbitrary point $p\in M$, a local coordinate $(x^1,\ldots,x^m)$, and $\displaystyle \left\{ \frac{\partial}{ \partial x^i} \right\}$ as the independent basis for $T_p M$. Then the linear map $\kappa: T_p M \to T_{U(p)}U(M)$, where for each $x^j$, $\kappa$ maps $\displaystyle v^j \frac{\partial}{\partial x^j}(p) \mapsto \exp\Big( L( p ) \Big) \sum_{i=1}^{n} v^j \frac{\partial f}{\partial x^j}(p) L_i$, yields $\kappa(T_p M) \subseteq T_{U(p)}U(M)$. 

    Now take $\hat{U}(p) \in T_{U(p)}U(M)$. By Lemma \ref{lemma:commutative-lie-algebra}, $\displaystyle \hat{U}(p) =  \exp\Big( L( p ) \Big) \sum_{i=1}^{n} \sum_{j=1}^{m} v^j \frac{\partial f_i}{\partial x^j}(p) L_i$. Thus, defining a map $\nu :T_{U(p)}U(M) \to T_p M$, $\displaystyle \exp\Big( L( p ) \Big) \sum_{i=1}^{n} \sum_{j=1}^{m} w^j \frac{\partial f}{\partial x^j}(p)L_i \mapsto \sum_{j=1}^{m} w^j \frac{\partial }{\partial x^j}$ yields $\nu\Big( T_{U(p)}U(M) \Big) \subseteq T_pM$. 
\end{proof}

\begin{remark}
    Goh et al. \cite{goh2023lie}, which consider the dynamical Lie algebra generated from the selected gates of a variational quantum circuit. The authors note that if roots that generate the circuit commute then the circuit is quite easy to approximate and simulate classically. Interestingly, we added to the authors work by noting the local equivalence between tangent vector space. 
\end{remark}

However, one may then ask whether to consider the independent basis as either the subset of root matrices or the independent coordinate basis of the data manifold. The general answer is the set of root matrices since the generation of the tangent vector space comes from this set. However, as will be displayed in Section \ref{subsec:vector-fields}, there are instances when one needs to consider the coordinate vector field. Section \ref{subsec:vector-fields} discusses these subtleties of vector fields and applications to curvature.

We now move on to the more general case of noncommutative matrices and the Lie algebra induced from this vector space. As previously noted, it is posited that this Lie algebra contains latent information about the vector space, but this is not shown. 

\subsection{Noncommutative Feature Maps}\label{subsec:noncommutative-tangent}

For the general case of Equation \eqref{eq:faeture-map}, this section discusses the structure of this derivative with respect to the underlying Lie algebra of the function $L(\cdot)$. From the known roots of the underlying Lie algebra, we may calculate the root matrices that generate the infinite sum in \eqref{eq:derivative-feature-map}. Once this is known, one may then derive a closed form of the Lie algebra generated by the tangent space. This Lie algebra is given to further generalize of the results of tangent space, with the assumption that dynamic Lie algebra will yield insight in future research.  

Recall that the encoding scheme $L(p)$ is the sum of skew-Hermitian matrices of dimension $2^{N}$. Ergo, $L(p) \in \mathfrak{su}(2^N)$. For the finite set of roots of $\mathfrak{su}(2^N)$, $\{ E_i \}$ there exists a subset of roots $\{E_{i'}\}$ that underlay the sum $L(\cdot)$. Of course, this might not be the final set of roots that generator the formula of the derivative. So, for $E_g, E_h \in \{E_{i'}\}$ if there exists a root matrix $E_k$ such that $[E_g,E_h] = c_{gh}E_k$ and $E_k \not\in \{E_i\}$ then append this set with $E_k$. Keep repeating until all roots are identified. Since $\{ E_d \}$ is finite, $\{E_{i'}\}$ is finite and the procedure will converge. Thus, the sum $\mathcal{L}(\cdot)$ can be decomposed into a sum of $\{E_{i'}\}$. Ergo, $\mathcal{L}(\gamma)(0)$ is a finite sum of root matrices with respective scalars that come from convergent and well-defined infinite sums;  one may see convergence from the exponential Taylor series, coupled with the smooth properties of each function $f_k$. 

With the logic above, we rewrite Equation \eqref{eq:derivative-feature-map} as 
\begin{equation} \label{eq:derivative-general-simplified}
    \frac{d}{dt} \exp\Big( L( \gamma(t) ) \Big) \bigg|_{t=0} = \exp\Big( L( \gamma(t) ) \Big) \sum_{i} \hat{f}_{i} E_{i}.
\end{equation}
Using this simplified form, for two tangent vectors $\displaystyle e^{L(p)} \sum_{i=1}^{n_e}  g_i E_i, \  e^{L(p)} \sum_{j=1}^{n_e} h_j E_j \in T_{U(p)}U(M)$, it is clear the commutator applied to these vectors has the form 
\begin{equation}\label{eq:general-commutator}
    \begin{split}
           & \left[ e^{L(p)} \sum_{i=1}^{n_e}  g_i E_i , e^{L(p)} \sum_{j=1}^{n_e} h_j E_j  \right] \\
           & = e^{L(p)} 
           \left(\begin{split}
             & \sum_{j=1}^{n_e}  g_{i} E_{i} e^{L(p)} \sum_{j} h_{j} E_{j} \\
             & - \sum_{j=1}^{n_e} h_j E_j e^{L(p)} \sum_{i} g_i E_i  
           \end{split} \right).
    \end{split}
\end{equation}
However, this closed from is still a bit cumbersome. Leveraging the linear property of the commutator, we will consider only single roots for the analysis to derive a closed form for this dynamic Lie algebra we are deriving.

\begin{remark}\label{remark:deriv-commutativity}
    Before utilizing Equation \eqref{eq:general-commutator}, recall that $(e^{L(p)})^{\dagger} = e^{-L(p)}$, and observe that the Lie derivative of $e^{L(p)}(e^{L(p)})^{\dagger} = \mathbb{I}$, where, from the commutativity of the operators, the Leibniz rule yields $\displaystyle e^{L(p)} \mathcal{L}(\gamma)(0)(e^{L(p)})^{\dagger} = \mathcal{L}(\gamma)(0)$. Using the same logic, it is also true that $\displaystyle (e^{L(p)})^{\dagger}\mathcal{L}(\gamma)(0)e^{L(p)} = \mathcal{L}(\gamma)(0)$.  
\end{remark}  

With the equalities noted in Remark \ref{remark:deriv-commutativity} and starting with two initial tangent vectors, each with one root matrix for simplicity of the analysis, one may confirm that
\begin{equation*}
\begin{split}
    & \Big[ e^{L(p)}E_i,e^{L(p)}E_j \Big]  = e^{L(p)} \Big( E_ie^{L(p)}E_j - E_j e^{L(p)}E_i \Big) \\
    & = \big( e^{L(p)} \big)^2  \Big( \big( e^{L(p)} \big)^{\dagger} E_ie^{L(p)}E_j - \big( e^{L(p)} \big)^{\dagger} E_j e^{L(p)}E_i \Big) \\
    & = \big( e^{L(p)} \big)^2 \Big( E_i E_j - E_j E_i \Big) = (e^{L(p)})^2 c_{ij} E_{k} .
\end{split}  
\end{equation*}

For general $m,n \in \mathbb{N}$ and two vectors in the ``semi-closed Lie subalgebra" generated by the tangent vector space, one may further verify 
\begin{equation*}
\begin{aligned}
    & \Big[\big( e^{L(p)} \big)^m E_i, \big( e^{L(p)} \big)^n E_j \Big]  \\
    & = \big( e^{L(p)} \big)^m E_i \big( e^{L(p)} \big)^n E_j - (e^{L(p)})^n E_j \big( e^{L(p)} \big)^m E_i \\
\end{aligned}  
\end{equation*}
\begin{equation*}
\begin{aligned}
    & = \big( e^{L(p)} \big)^{m+n} \left( 
    \begin{aligned}
       &  \big( (e^{L(p)} )^{n} \big)^{\dagger}E_i \big( e^{L(p)} \big)^n E_j \\
       & -  \big( (e^{L(p)} )^{m} \big)^{\dagger} E_j \big( e^{L(p)} \big)^m E_i 
    \end{aligned} \right) \\
    & = (e^{L(p)})^{m+n} \left( 
    \begin{aligned}
       & \big( (e^{L(p)})^{n} \big)^{\dagger}E_i (e^{L(p)})^n E_j \\
       & - \big( (e^{L(p)})^{m} \big)^{\dagger} E_j (e^{L(p)})^m E_i 
    \end{aligned} \right) \\
    & = \big( e^{L(p)} \big)^{m+n} \Big( E_i E_j - E_j E_i \Big) = \big( e^{L(p)} \big)^{m+n} c_{ij} E_{k} . 
\end{aligned}  
\end{equation*}
Therefore, an element in the Lie algebra will be a finite sum of the form $\displaystyle \sum_{k=1}^{d} \big( e^{L(p)} \big)^{n_k} c_k L_{k}$. Using the flexible notation of direct sum, we have that the Lie algebra generated by the vector tangent space has the form 
\begin{equation*}
     \mathfrak{U}_p = \bigoplus_{l=1}^{\infty} \big( e^{L(p)} \big)^l \mathfrak{su}_{\mathcal{L}(p)}.  
\end{equation*}

Collecting the arguments within this subsection brings us to the following proposition.
\begin{proposition}\label{prop:closed-form-algebra}
    Given a feature map of the form in Equation \eqref{eq:faeture-map}, a data manifold $M$, and a point $p \in M$ the Lie algebra generated from the tangent vector space on the subspace of $\SU(N)$ has the general form
\begin{equation}\label{eq:general-tangent-Lie-algebra}
     \mathfrak{U}_p = \bigoplus_{l=1}^{\infty} \big( e^{L(p)} \big)^l \mathfrak{su}_{\mathcal{L}(p)}.  
\end{equation}
\end{proposition}

We now have the necessary foundation to establish a Riemannian structure on this space, and, for the purposes of generality, we establish this space from the dynamic Lie algebra in Equation \eqref{eq:general-tangent-Lie-algebra}. However, this is a lofty task that has the potential of deviating from the focus of the tangent tangent space. Consequently, a simplification of the structure of an element is needed, at least, to establish a Riemannian geometry where an arbitrary tangent space is a particular case. Thus, we require the following assumption.

\begin{assumption}\label{ass:same-powers}
For elements $X_,X_2,\ldots, X_n \in \mathfrak{U}_p$ there is a positive natural number $l$ such that for all $i$, $\displaystyle X_i = \big( e^{L(p)} \big)^l L_{X_i}$, where $L_{X_i}$ is the sum of skew-Hermitian operators with respect to $X_i$. 
\end{assumption}




\section{Riemannian Metric and Levi-Civita Connection}\label{sec:geodesics}

Now that we have established an analytic form of a tangent bundle, we move on to a closed form to calculate the geodesics of points in the range. This entails a Riemannian metric and the `shortest' path between points; for some background, see Petersen \cite{petersen2006riemannian} or Gallier and Quaintance \cite{gallier2020differential}. With interest in expanding this mathematical framework to a general Lie groups, the derivations and respective proofs will consider the Lie algebra generated from the vector tangent space.  

\begin{remark}
 Note that the set of functions $\{f_k\}$ stay fixed since these functions were selected a priori and are essential to the encoding scheme. Furthermore, the derivation of the tangent space at an arbitrary point $p\in M$ is contingent on the equality $\displaystyle (e^{L(p)})^{\dagger}\mathcal{L}(\gamma)(0)e^{L(p)} = \mathcal{L}(\gamma)(0)$, which does not hold in general for another set of functions $\{f'_i\}$ as $e^{L(p)}(e^{L'(p)})^{\dagger} \neq \mathbb{I}$, for $\displaystyle L'(p) = \sum_{i=1}^{n}f'_i(p) L_i$. Thus, the equality does not hold for the entirety of the tangent bundle.     
\end{remark}

For Lie groups one typically applies the Killing form (or Cartan-Killing metric), $B(H,L) := \Tr(ad_H \circ ad_K)$, where $ad_H(E) = [H,E]$ and $H,K,E \in \mathfrak{g}$. Interestingly, the Killing form can be used to identify $\mathfrak{g}$ with its dual $\mathfrak{g}^{*}$; see Gracia-Bond{\'\i}a, V{\'a}rilly, and Figueroa \cite{gracia2013elements}. The Killing form on $\mathfrak{su}(N)$ has the closed form $B(H,K) = 2N\Tr(HK)$ and is derived by using the roots of $\mathfrak{su}(N)$; see Gallier and Quaintance \cite{gallier2020differential} and Jost \cite{jost2008riemannian}. Furthermore, as noted in Lewis et al. \cite{lewis2024geodesic}, the function
\begin{equation}\label{eq:herm-metric}
 g_{\mathfrak{su}}(H,K) = \Tr(H^{\dagger}K)/N   
\end{equation}
applied to $\mathfrak{su}(N)$ is a Riemannian metric on $T_e\SU(N)$. Since $T_{U(p)}U(M) \subset T_{U(p)}\SU(N)$, one would expect this metric to also apply. However, as we will show, this metric is incomplete and requires adjustment that satisfies the definition of a metric applied to our tangent space $T_{U(p)}U(M)$. 

Applying Equation \eqref{eq:herm-metric} to arbitrary vectors in the dynamic Lie algebra in Proposition \ref{prop:closed-form-algebra} may yield a complex value. However, given the properties of the trace, we claim for $\displaystyle H,K \in \bigoplus_{l=1}^{\infty} \big( e^{L(p)} \big)^l \mathfrak{su}_{\mathcal{L}(p)}$, $g_{\mathfrak{su}}(H,H) \in \mathbb{R}$ and $g_{\mathfrak{su}}(H,K) = \overline{g_{\mathfrak{su}}(K,H)}$. This is shown below. 

\begin{claim}\label{claim:pos-conjugate-metric}
Given a smooth Riemannian manifold $M$ and a feature map of the form in Equation \eqref{eq:faeture-map}, for $\displaystyle H,K \in \bigoplus_{l=1}^{\infty} \big( e^{L(p)} \big)^l \mathfrak{su}_{\mathcal{L}(p)}$, we have that $g_{\mathfrak{su}}(H,H) \in \mathbb{R}$, and $g_{\mathfrak{su}}(H,K) = \overline{g_{\mathfrak{su}}(K,H)}$.
\end{claim}
\begin{proof}
 Take $\displaystyle H = \sum_{i=1}^{\beta}\big( e^{L(p)} \big)^{l_i} H_i$ and $\displaystyle K = \sum_{j=1}^{\alpha}\big( e^{L(p)} \big)^{w_j} K_j$. Showing the real property, observe 
\begin{equation*}
    \begin{split}
     & N\cdot g_{\mathfrak{su}}(H,H) \\
     &= -\sum_{i}^{\beta}\Tr(-H_i H_i) \\
    & + \sum_{i=1}^{\beta}\sum_{i \neq j}^{\beta}\Tr \Big( -H_i e^{L(p)} \big)^{l_j - l_i} H_j \Big) \\
    & = \sum_{i}^{\beta}-\Tr(H_i H_i) \\
    & + \sum_{i=1}^{\beta}\sum_{i<j}^{\beta}
    \left(
    \begin{split}
       &     \Tr \Big( -H_i\big( e^{L(p)} \big)^{l_j - l_i} H_j \Big)  \\
       & + \Tr \Big( -H_j \big( e^{L(p)} \big)^{l_i - l_j} H_i \Big)
    \end{split}
    \right) \\
    & = \sum_{i}^{\beta}-\Tr(H_i H_i) \\
    & + \sum_{i=1}^{\beta}\sum_{i<j}^{\beta}
    \left(
    \begin{split}
       &     \Tr \Big( -H_i\big( e^{L(p)} \big)^{l_j - l_i} H_j \Big)  \\
       & + \overline{\Tr \Big( H_i\big( e^{L(p)} \big)^{l_j - l_i}\big)^{\dagger} H_j \Big)}
    \end{split}
    \right).
    \end{split}
\end{equation*}
Therefore, $g_{\mathfrak{su}}(H,H)$ is real. It is clear that when $H$ is a term with a single $(e^{L(p)})^l$ then $g_{\mathfrak{su}}(H,H)$ is positive. 

The second property follows from the observation in the equality above. 
\end{proof} 

\begin{remark}
    For $\displaystyle \displaystyle H = \sum_{i=1}^{\beta}\big( e^{L(p)} \big)^{l_i} H_i$ one may derive the equality
\begin{equation*}
    \begin{split}
         g_{\mathfrak{su}}(H,H)  = & -\sum_{i=1}^{\beta} \Tr\Big( H_i H_i \big) \\ &
         -2 \sum_{j=2}^{\beta} \sum_{i<j} \mbox{Re}\Big(\Tr \Big( H_j H_i \big( e^{L(p)} \big)^{l_j - l_i} \Big) \Big),
    \end{split}
\end{equation*}
which displays that $g_{\mathfrak{su}}$ applied to a tangent vector yields a sum of metrics on $\mathfrak{su}(N)$, along with `residual' terms. This shows the complexity that the vector space a quantum feature map yields compared to the simple view of starting with $\mathfrak{su}(N)$ at the identity matrix with a left invariant metric. 
\end{remark}

One may see that $g_{\mathfrak{su}}$ is left and right invariant against multiplication of the vectors by $\big( e^{L(p)} \big)^n$, for $n\in\mathbb{N}$. Therefore, the function
\begin{equation}\label{eq:final-metric}
 g(H,L) := \frac{1}{2N} \Big( \Tr(H^{\dagger}L)  + \Tr(L^{\dagger}H) \Big)
\end{equation}
holds the properties of the definition of a Riemannian metric. This is the metric that will be applied to the tangent vector spaces. As is typical in the literature of Riemannian geometry, the notation of $g(\cdot,\cdot)$ and $\langle \cdot, \cdot \rangle$ are interchangeable, and will be used when there is no confusion. For computation, notice that $g(H,L)$ is to just the real part of $g_{\mathfrak{su}}(H,L)$. 

We may now discuss geodesics. At first thought, it is intuitive for a geodesic on $M$ to generate a geodesic on $U(M)$. However, this is only an ansatz. Ergo, a closed form geodesic on $U(M)$ is not clear. The following proposition shows that geodesics on $M$ and $U(M)$ are bijective, thus simplifying the computation of distances on $U(M)$. Observe that the following proposition uses the fact that metric given in Equation \eqref{eq:final-metric} multiplied by $1/2$ is a Levi-Civita connection. Although not yet discussed, Levi-Civita is described and justified in Section \ref{subsec:curvature}, and only used to quickly show one-direction.

\begin{proposition}\label{prop:geodesic-form}
    For a quantum feature map that holds Assumption \ref{ass:general-map}, every geodesic on $U(M)$, $\gamma_U:[a,b] \to U(M)$, has the form $\gamma_U(\cdot) = \exp\Big( L\big(\gamma(\cdot) \big) \Big)$ for $\gamma(\cdot)$ a geodesic on $M$.
\end{proposition} 
\begin{proof}
From the form of the Levi-Civita connection in Equation \eqref{eq:def-levi-civita}, it is clear that any geodesic $\gamma(\cdot)$ on the data manifold is also a geodesic on $U(M)$.

Now take $\gamma_U:[a,b] \to U(M)$ as a geodesic on $U(M)$. We will show that when two operators on the geodesic are close then each scalar-valued function $f_k$ in the encoding scheme are extremely close. Thus, one may then infer the existence of a geodesic on the data manifold driving the geodesic on $U(M)$. 

Denote $d_{U(M)}$ as the distance function on $U(M)$. By assumption, for an arbitrary $t\in(a,b)$ there exists a neighborhood around $t$, call it $B_t$, such that $d_{U(M)}( \gamma_U(t_1), \gamma_U(t_2) ) = \nu |t_2 - t_1|$, for some positive constant $\nu > 0$ and any $t_1,t_2 \in B_t$. Fixing $t_1,t_2 \in B_t$, observe that there exists points $x_1,x_2 \in M$ such that $\gamma_U(t_i) = \exp\big( L(x_i) \big)$ for $i=1,2$. Now, denoting $d_{SU}$ as the distance function on $\SU(N)$ for an arbitrary dimension $N$, we have that $d_{SU} = || \log(U_1^{\dagger} U_2) ||_{F}$ where $U_1, U_2 \in \SU(N)$ and $|| \cdot ||_{F}$ is the Frobenius norm on matrices; see Jost or Pertici and Dolcetti \cite{jost2008riemannian,pertici2024some} for further information. Since $U(M) \subset \SU(N)$, we have $d_{U(M)}( \gamma_U(t_1), \gamma_U(t_2) ) \geq  d_{SU}( \gamma_U(t_1), \gamma_U(t_2) )$ for all $t_1,t_2 \in B_t$. Furthermore, since the manifold is embedded, $d_{M}(\cdot,\cdot) \leq L^2(\cdot, \cdot)$. Now, using the two inequalities above, the properties of the functions $f_k$, the fact that $\mbox{det}(U_i) = 1$ for all $i$, and the BCH formula, we have

\begin{equation*}
    \begin{split}
       & d_M( x_2, x_1 ) \leq \beta_1 \cdot L^2(x_2 ,x_1) \\
       & \leq \beta_1 \cdot \sqrt{ \sum_{k=1}^{k} |f_k(x_2) - f_k(x_1)|  } \\
        & \leq \beta_1 \cdot \sqrt{ \sum_{k=1}^{n} \sum_{i,j = 1}^{N} |f_k(x_2) - f_k(x_1)|\cdot | U_k^{ij} | } \\
        & = \beta_1 \cdot \big\| L(x_1)^{\dagger} + L(x_2) \big\|_{F} \\
        & = \beta_1 \cdot \big\| \log \Big(  \exp\big(  L(x_1)^{\dagger} + L(x_2) \big) \Big) \big\|_{F} \\
        & \leq \beta_1 \cdot \big\| \log \big( \gamma_U(t_1)^{\dagger} \gamma_U(t_2) \big) \big\|_{F}  \\
        & \leq \beta_1 \cdot d_{U(M)} \big( \gamma_U(t_1), \gamma_U(t_2) \big) = \nu \cdot \beta_1 \cdot  |t_2-t_1|. 
    \end{split}
\end{equation*}

Combining the inequality above, taking the smooth property of the $f_k$ functions and the local compactness of $M$, there exists a finite constant $\alpha > 0$ where $d_M(x_1,x_2) \leq \alpha \cdot |t_2 - t_1|$ for all $t_1,t_2 \in B_t$. Thus, for an arbitrarily small $\epsilon>0$, there exists a finite increasing set of times $0=t_0<t_1<\cdots<t_{n_{\epsilon}} = 1$ and a finite set $\{v_0,\ldots,v_{n_{\epsilon}}\}$ of points on the data manifold $M$ where $v_i \mapsto \gamma_U(t_i)$, then $d_M( v_i, v_{i+1} )<\epsilon$ for all $i<n$. Therefore, there exists a geodesic on $M$ that maps the points in $U(M)$.  
\end{proof}

\begin{remark}
Now that we have a closed form for a geodesic on $U(M)$, say $\gamma_U:[a,b] \to U(M)$, one may now answer a side question: for approximation purposes, what is the farthest away a special unitary operator is from a geodesic? Specifically, for a given $\epsilon > 0$, can we solve
\begin{equation*}
    \sup_{P \in SU(N)} \inf_{\epsilon_0 \leq \epsilon } \Big\{  \gamma_U\big( [a,b] \big) \cap B(P,\epsilon_0) \neq \emptyset  \Big\},
\end{equation*}
where $B(P,\epsilon)$ is a ball of size $\epsilon$ with respect to the distance function $d_{SU}$ in Proposition \ref{prop:geodesic-form}. Answering this question gives a geometric way to determine how `close' an approximation is to the base operator. 

From the BCH formula, for $P \in \SU(N)$ take $\Tilde{P} \in \mathfrak{su}(N)$ such that $P = e^{\Tilde{P}}$, with the distance function on $\SU(N)$ we see that 

\begin{equation}\label{eq:closet-distance-bch}\medmath{
    \begin{aligned}
        & \min_{t\in[a,b]} \Big\| \log\big( \big( e^{L(\gamma(t)) }\big)^{\dagger} e^{\Tilde{P}} \big) \Big\|_F  \\
         = & \min_{t\in[a,b]} \left\| \begin{aligned}
        &e^{L(\gamma(t))^{\dagger}} + \Tilde{P} +\frac{1}{2}\big[ e^{L(\gamma(t)^{\dagger}},\Tilde{P} \big] \\
        & + \frac{1}{12}\Big( \big[ e^{L(\gamma(t))^{\dagger}},\big[e^{L(\gamma(t))^{\dagger}} ,\Tilde{P} \big]\big] + \big[ \Tilde{P},\big[ \Tilde{P},e^{L(\gamma(t))^{\dagger}} \big]\big] \Big) \\
        & + \frac{1}{24}\big[\Tilde{P},\big[e^{L(\gamma(t))^{\dagger}},\big[e^{L(\gamma(t))^{\dagger}} ,\Tilde{P} \big] \big] \big] + \ldots 
        \end{aligned} 
        \right\|_F 
    \end{aligned} }
\end{equation}
finding an analytic form is quite difficult. However, in general, an approximation of the minimum distance is straight forward to calculate.   
\end{remark}

We now move our attention to deriving closed form expressions to calculate curvature. 

\subsection{Calculation of Curvature}\label{subsec:curvature}

The previous subsections established the necessary properties to give explicit formulas to calculate the curvature of our space $U(M)$, the range of a feature map given in Equation \eqref{eq:faeture-map}. To calculate curvature, many of the results from the literature concern that of a \define{bi-invariant metric} of a Lie group. Hence, for a Lie group $G$ and any $h \in G$, both maps $l \mapsto h \cdot l$ or  $l \mapsto l \cdot h$ are isometric, which are respectively \define{left-invariant} and \define{right-invariant}. Of course, this definition makes sense as the multiplication heavily uses the group structure. As noted above, this the group structure does not hold with a general quantum feature map of the form in Equation \eqref{eq:faeture-map}. Consequently, closed form terms that calculate curvature require a ground-up approach to establish validity. Using results in the previous sections, we will show that a similar structure of curvature holds. 

Strictly from the definition, the \define{adjoint action} or adjoint representation of a Lie algebra has the form
\begin{equation}\label{eq:adjoint}
    (\mathrm{ad}X)Y = [X,Y].
\end{equation}
Clearly, $\mathrm{ad}X$ is a linear transformation from $\mathfrak{U}_p$ to itself. Using the properties of the trace, one may see that $\Big\langle (\mathrm{ad}X)Y , Z \Big\rangle = \Big\langle Y , (\mathrm{ad}X)^{\dagger}Z \Big\rangle$. However, in general, it is not clear that $\Big\langle (\mathrm{ad}X)Y , Z \Big\rangle = -\Big\langle Y , (\mathrm{ad}X)Z \Big\rangle$, ergo, $(\mathrm{ad}X)$ is skew-symmetric. The claim below proves this property. 

\begin{claim}\label{claim:skew-symm}
    For $X,Y,Z \in \mathfrak{U}_p$ that hold Assumption \ref{ass:same-powers}, then $\left\langle (\mathrm{ad}X)Y , Z \right\rangle = -\left\langle Y , (\mathrm{ad}X)Z \right\rangle$.
\end{claim}
The proof of the claim uses the properties of the trace and equalities in Remark \ref{remark:deriv-commutativity}. Since the proof is simple analysis that does not yield any insight, it is not included for brevity. Moreover, while the claim is fairly restrictive, Assumption \ref{ass:same-powers} holds for vector fields.   

For completeness, we state the Killing form. The general definition of the \define{Killing form} is given by 
\begin{equation}\label{eq:killing}
    (X,Y) := \Tr(\mathrm{ad}X \circ \mathrm{ad}Y),
\end{equation}
and is calculated by using the root matrices. See Renteln \cite{renteln2014manifolds} or Jost \cite{jost2008riemannian} for more information on the adjoint action and Killing form. 

From here we focus on \define{affine connections} on the tangent bundle, which assigns a smooth vector fields from two given smooth vector fields. This is also called the \define{covariant derivative}, where for smooth vector fields $X$ and $Y$, $\nabla_X Y$ is defined to be the derivative of $Y$ in the direction of $X$. There is a special connection, called the \define{Levi-Civita connection}, that is metric compatible and torsion-free. We claim that Levi-Civita connection, which we fix the notation of $\nabla_X Y$, is given by the formula
\begin{equation}\label{eq:def-levi-civita}
 \nabla_{X}Y = \frac{1}{2} (\mathrm{ad}X)Y = \frac{1}{2}[X,Y].
\end{equation}
As noted on page 237 in Renteln \cite{renteln2014manifolds}, this is precisely the connection for a left invariant Riemannian metric on $\SU(N)$ at the identity. However, given the restriction of $\SU(N)$ to the space $U(M)$, we will show this holds by applying Theorem 8.4 in Renteln \cite{renteln2014manifolds}. It is clear that all assumptions in the theorem hold for $\nabla_{X}Y$, except for the last assumption of metric compatibility. For brevity, these details are omitted. The metric compatibility is shown in the claim below. 

Using standard notation, for a smooth vector field $X$ and differentiable function $f:M\to\mathbb{R}$, denote the derivative in the direction of $X$ as $Xf$. There are subtleties with respect to the directional derivative of a vector field as the derivative needs to be in the same vector space. However, these subtleties are not given in this manuscript and the references of Jost and Petersen \cite{jost2008riemannian,petersen2006riemannian} are suggested for further reading.   

\begin{claim}\label{claim:metric-compat}
    For smooth vector fields $X,Y,Z \in \Gamma(TU(M))$ and for $\nabla_{X}Y$ given by Equation \eqref{eq:def-levi-civita}, it holds that $Xg\big(X,Y\big) = 0 = g\big(\nabla_{X}Y,Z\big) + g\big(Y,\nabla_{X}Z\big).$
\end{claim}
\begin{proof}
We first show that $g\big(\nabla_{X}Y,Z\big) + g\big(Y,\nabla_{X}Z\big) = 0$. Denote $Y = e^{} \bar{Y}$, and recall from Remark \ref{remark:deriv-commutativity} that $e^{L}$ and $e^{-L}$ is commutative with $\bar{Y}$, with the understanding that the vector fields are localized to each data point. Using the shifting properties of the trace and the remark, taking an arbitrary $p\in M$ we see that
\begin{equation*}
    \begin{split}
        & g\big(\nabla_{X}Y,Z\big) =\Tr \big( (XY-YX)^{\dagger} Z \big) + \Tr \big( (XY-YX) Z^{\dagger} \big) \\
        & = \Tr\big(Y^{\dagger}X^{\dagger}Z - X^{\dagger}Y^{\dagger}Z\big) + \Tr\big( Z^{\dagger}XY - Z^{\dagger}YX \big) \\
        & = \Tr\big( \bar{X} e^{-L(p)} \bar{Z} \bar{Y}- \bar{Z}e^{-L(p)}\bar{X}\bar{Y}\big) \\
        & \hspace{3pt}  + \Tr\big( \bar{Y} \bar{Z} e^{L(p)} \bar{X} - \bar{Y} \bar{X}e^{L(p)}\bar{Z}\big) = -g\big(Y,\nabla_{X}Z\big). 
    \end{split}
\end{equation*}

For the LHS, the equality is shown by taking the derivative at $p$, taking the limit inside the trace, which is possible since the trace is linear and continuous, then applying the well-known equality of the Lie derivative $\mathfrak{L}$ on smooth vector fields, $\mathfrak{L}_{X}Y = [X,Y]$, and the Leibnitz rule; see Petersen \cite{petersen2006riemannian}. Typically, the Lie derivative is denoted as $\mathcal{L}$ which is why this notation has not been adjusted as $\mathcal{L}$ already defined. Writing this mathematically, take an arbitrary $p \in M$ and $\phi_p(t)$ as the path for $X$, then 
\begin{equation*}
    \begin{split}
        & Xg\big(X,Y\big) = \lim_{t \to 0} \frac{\phi^{-1}\big( g\big(X,Y\big)|_{\phi_p(t)} \big) - g\big(X,Y\big)|_{p}}{t} \\
        & = \Tr\big( [X,Y^{\dagger}]Z + Y^{\dagger}[X,Z] \big) \\
        & + \Tr\big( [X,Z^{\dagger}]Y + Z^{\dagger} [X,Y] \big) \\
        & = \Tr\big(  Y^{\dagger}[X,Z] \big)  + \Tr\big( [X,Z^{\dagger}]Y \big) \\
        & + \Tr\big( [X,Y^{\dagger}]Z)  + \Tr\big(Z^{\dagger} [X,Y] \big) \\
        & = \Tr\big(  Y^{\dagger}[X,Z] \big)  - \Tr\big( Y^{\dagger} [X,Z] \big) \\
        & + \Tr\big( Z[X,Y^{\dagger}])  - \Tr\big(Z [X,Y^{\dagger}] \big) = 0.
    \end{split}
\end{equation*}
The second to last equality comes from following the logic showing the LHS.
\end{proof}

Observe in the proof that objects are vector fields and not arbitrary elements in the Lie algebra $\mathfrak{U}_p$, for an arbitrary $p\in M$. However, the proof still holds for elements in $\mathfrak{U}_p$ that all hold Assumption \ref{ass:same-powers}. Moreover, the condition that $g\big(\nabla_{X}Y,Z\big) + g\big(Y,\nabla_{X}Z\big) = 0$ is the same assumption in Milnor \cite{milnor1976curvatures} that is used to establish many results.  

For the first curvature, we consider \define{curvature tensor}, 
\begin{equation}
    R_{XY} : = \nabla_{[X,Y]} - \nabla_Y \nabla_X + \nabla_X \nabla_Y.
\end{equation}
Curvature tensor measures the commutativity of the covariant derivative, displaying how much the manifold deviates from a flat structure. Taking Claim \ref{claim:skew-symm} and following the logic in Milnor \cite{milnor1976curvatures}, we may simplify the curvature tensor to  
\begin{equation}
    R_{XY} = \frac{1}{4}\mathrm{ad}[X,Y],
\end{equation}
considerably simplifying the computation of smooth vector fields. 

\begin{remark} \label{remark:milnor}
While the logic in Milnor \cite{milnor1976curvatures} does take the assumption of left-invariance, the skew-symmetric property and equalities derived in Milnor are all that is required to derive analogous results. 
\end{remark}

From here we may define the \define{sectional curvature}, 
\begin{equation}\label{eq:sec-curve}
    \kappa(X,Y) := \frac{\left\langle R_{XY}(X),Y \right\rangle}{\left\langle X,X \right\rangle\left\langle Y,Y \right\rangle - \left\langle X,Y \right\rangle^2},
\end{equation}
where $X,Y \in T_{U(p)}U(M)$ are linearly independent with respect to the metric $g(\cdot, \cdot)$. The terms in the denominator normalize the calculation. The function $\kappa(X,Y)$ measures how a two-dimensional surface, or section, inside the tangent space curves. When $X$ and $Y$ are orthonormal we may simplify the computation of the sectional curvature to
\begin{equation}\label{eq:sec-curve-ortho}
    \kappa(X,Y) = \frac{1}{4} \Big\langle [X,Y], [X,Y] \Big\rangle.
\end{equation}

From the sectional curvature, we define the \define{Ricci curvature} which measures the distortion of the volume, and can be thought of as the Laplacian of the Riemannian metric. As given in Petersen \cite{petersen2006riemannian}, take $O_1, \ldots, O_n \in T_{U(p)}U(M)$ to be an orthonormal basis with respect to the metric $g(\cdot, \cdot)$, then for $X,Y \in T_{U(p)}U(M)$
\begin{equation}\label{eq:ricci-curv}
    Ric(X,Y) = \Tr\big( Z \to R_{ZX}Y \big) = \sum_{i=1}^{n} \Big\langle R_{O_iX}(Y),O_i \Big\rangle.
\end{equation}

Finally, we take the definition of \define{scalar curvature} from Petersen \cite{petersen2006riemannian}, where this curvature gives an aggregated number of localized curvature. The definition states
\begin{equation}\label{eq:scalar-curvature}
    \mbox{scal}:= \Tr(Ric) = 2\sum_{i<j} \kappa(O_i,O_j),
\end{equation}
where the $O_i$'s are an orthonormal basis previously noted. Observe that there are other forms for the scalar curvature, but given the computational ease of sectional curvature, these equalities are omitted.  

While the closed forms given allow for ease of computation, there is the question of how to compute the orthonormal basis of a vector field and if the focus should be on the tangent vector field of the domain or the roots of the Lie algebra that generate $\mathfrak{su}_{\mathcal{L}}$. The following subsection discusses these different perspectives. 

\subsection{\label{subsec:vector-fields} Different Perspectives of a Vector Field and Application of Curvature}

There is ambiguity of what is exactly meant by an orthonormal basis of a vector field in this setting since a tangent vector is with respect to both the coordinate system of $M$ and the root matrices that generates $\mathfrak{su}_{\mathcal{L}}$. If one considers a coordinate basis then the terms in the vector field are grouped with respect to the independent partials $\displaystyle  \left\{\frac{\partial}{\partial x^j} \right\}$, but only from the non-zero terms in the sum $\displaystyle \sum_{k = 1}^{n} \sum_{j=1}^{m} \frac{\partial f_{k}}{\partial x^j}\frac{\partial}{\partial x^j} L_{k} $. The application of the operator $\displaystyle \frac{\I - e^{-\mathrm{ad}_{L}}}{\mathrm{ad}_{L}}$ then yields the full form of a vector field. Of course, in general, there is no reason that these terms are orthonormal with respect to the metric given by Equation \eqref{eq:final-metric}. To get an orthonormal basis for the independent partials one may apply the Gram–Schmidt process. 

For the basis elements with respect to the root matrices, at first glance, this computation may seem purely mathematical. However, calculating the root matrices are tangible; in fact, for the worst case scenario of just given a skew-Hermitian matrix, Georges et al. \cite{georges2025pauli} state that for the size of the matrix $N$, the complexity is $O\big( N^2\log(N) \big)$ to decompose.  Considering that $\mathfrak{su}_{\mathcal{L}}$ is a Lie subalgebra of $\mathfrak{su}(2^N)$, then by the Cartan decomposition the root operators of $\mathfrak{su}_{\mathcal{L}}$ are essentially tensors of Pauli operators and control gates; see Varadarajan \cite{varadarajan2013lie}. Specifically, the root matrices are tensors of the identity matrix $\mathbb{I}$, the root operators of $\mathfrak{su}(2)$, and the root operators of $\mathfrak{su}(4)$. 

For computation we first need to derive all of the root matrices of the vector field. Starting with the operators $\{L_{k}\}$, for each $k$ represent the operator $L_{k}$ as tensor elements from roots in $\mathfrak{su}(2)$ and $\mathfrak{su}(4)$. Repeated application of the commutator will find the other root elements; see comments in Section \ref{subsec:noncommutative-tangent}. Once the root elements have been identified, the coefficients for each term needs calculated. This, again, can be accomplished through repeated recursive applications of the commutator given in Equation \eqref{eq:derivative-feature-map}. Of course, for applications, only a finite number of recursive terms are possible. Given how fast the denominator goes to infinity, a finite approximation will be sufficient. Since the root elements are already orthonormal, only individual coefficients need to be normalized.   

To final answer the answer of whether to use one orthonormal basis over another is given by the task. To illustrate this, we sample from the Poincar\'e half-plane model---see Jost \cite{jost2006compact}---and take the two encoding schemes of Angle~\cite{schuld21,schuld2019quantum} and IQP ~\cite{Shepherd_2009,bremner2016average,havlivcek2019supervised}. The Poincar\'e half-plane model is a hyperbolic space and has been shown to have uniform curvature. Particularly, for the two dimension case the Gaussian curvature is $\kappa = -1$; see Jost \cite{jost2008riemannian}, and Renteln \cite{renteln2014manifolds} for Gaussian curvature. For consistency of notation, the Poincar\'e half-plane model is denoted as $M$. However, given the size of the manifold, we restrict the real values for the pair $(x,y)$, where $x\in [-1,1]$ and $y \in [.1,1.1]$, and for computation, both intervals are discretized at increments of $.1$. 

For two points on $M$, $p_1 = (x_1,y_1)$ and $p_2 = (x_2,y_2)$, the distance between the points is either a line or an arc. Hence, 
\begin{equation}
    d(p_1,p_2) = 
    \left\{
    \begin{array}{l}
       |y_1 - y_2|   \mbox{ when } x_1 = x_2  \\
        2\cdot\log\left( \frac{||p_2 - p_1||_2 + ||p_2 - (x_1,-y_1)||_2 }{2\cdot \sqrt{y_1y_2}} \right)  \mbox{else}
    \end{array}
    \right. .
\end{equation}

\begin{figure*}[ht!]
    \centering
    \includegraphics[width=0.8\linewidth]{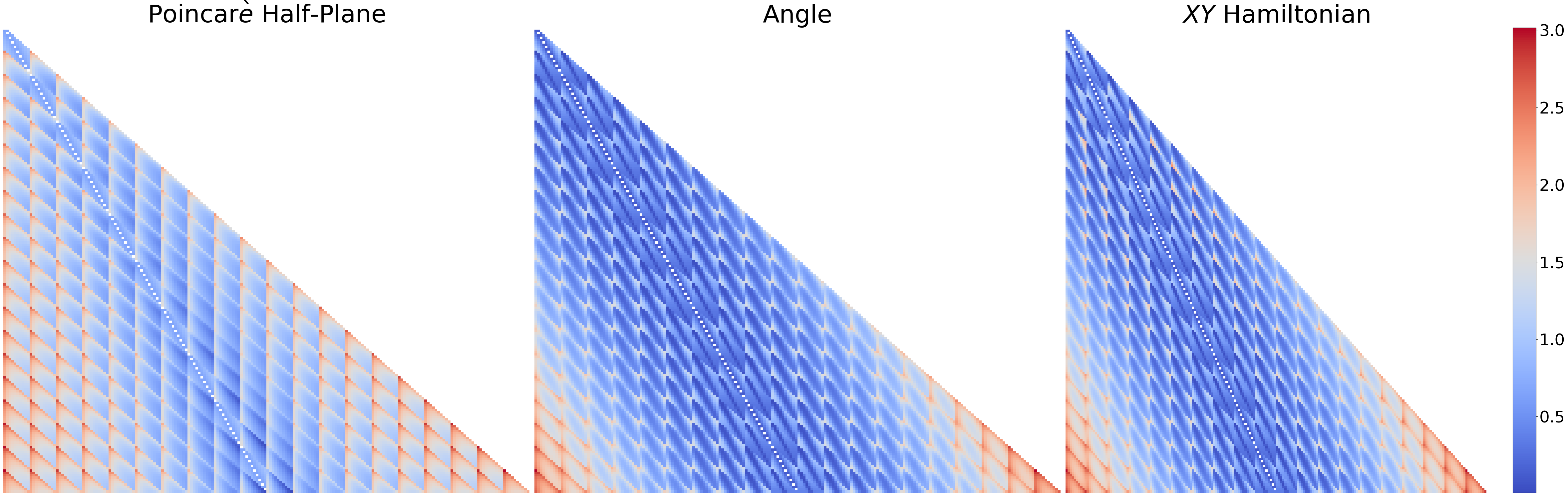}
    \caption{Heatmap of the distances between the points on the plane with respect to the metric, the Angle feature map, and the instantaneous quantum polynomial (IQP)  feature map.}
    \label{fig:distances}
\end{figure*}

For the encoding schemes, denote the Pauli operator $P \in\{X,Y,Z\}$ acting on the $i^{th}$ qubit as $\sigma^i_P$. Hence, $\displaystyle \sigma^i_P = \mathbb{I} \otimes \mathbb{I} \otimes \cdots \otimes P \otimes \mathbb{I} \cdots \otimes \mathbb{I}$. Considering the range of interval values in $M$, angle encoding has the form  $(x,y) \mapsto \exp\Big( -ix\cdot\sigma^1_X  -iy\cdot \sigma^2_X\Big)$, and the IQP encoding schemes follows the $XY$ Hamiltonian and is taken to be $(x,y) \mapsto \exp\Big( -ix\cdot\sigma^1_Y  -iy\cdot \sigma^2_Y -ixy \cdot X \otimes X\Big)$. For the distances, the $3^{rd}$ degree approximation of the derivative was used, and since angle has commutative roots, no approximation was necessary. Lastly, given Proposition \ref{prop:geodesic-form}, we directly map the geodesics on the Poincar\'e half-plane to the space of quantum operators. 

Figure \ref{fig:distances} displays the comparative distances between every point in the cloud. Interestingly, the small distance lines on the half-plane heatmap, which show the borders carved out from the lines, are exhibited in both encoding schemes. Importantly, the symmetric `perforated' line is evident in all heatmaps, indicating a high potential of similar curvature, at least, locally. 

For the curvature, one may see the angle feature map has the orthonormal elements $U(p)\big\{i \sigma^1_X \frac{\partial}{\partial x} ,i \sigma^2_X \frac{\partial}{\partial y} \big\}$ in the tangent bundle $TU(M)$. Of course, these elements are agnostic to either the root matrices or coordinate vector field. Applying the sectional curvature from Equation \eqref{eq:sec-curve-ortho} yields a flat curvature of $\kappa = 0$. In fact, in general, one may show that a quantum feature map in the form of Equation \eqref{eq:derivative-feature-map-commutative} will yield a flat structure. This is captured in the claim below.

\begin{claim}
    An encoding scheme in the form of Equation \eqref{eq:derivative-feature-map-commutative} will have flat curvature with respect to the Lie algebra root system, as well as when the independent coordinate basis has the mapping  $\{\frac{\partial}{ \partial x^{\alpha}}\} \to \{L_{k}\}$, where $L_{k}$ is in the sum $L(p)$ given in Equation \eqref{eq:sum-of-operators}. 
\end{claim}
\begin{proof}
    The fact that all of the elements and differentials are commutative yields the claim. 
\end{proof}

For the $XY$ Hamiltonian encoding scheme, we focus on the  orthonormal basis with respect to the coordinates. For this encoding scheme we follow the processes described in this subsection. But, for simplicity, we apply the first degree approximation. Applying the Gram-Schmidt process yields that the orthonormal basis elements are of the form
\begin{equation*}
    U(p) \left\{
    \begin{split}&  \Big(  - ip^2 \sigma_X \otimes \sigma_X -i \sigma^1_Y \Big)\frac{\partial}{ \partial p^1}, \\
    & \Big(- ip^1  \sigma_X \otimes \sigma_X - i \sigma^2_Y \Big)\frac{\partial}{ \partial p^2} \\
    &   - \frac{p^1p^2}{(p^2)^2 +1} \Big(  - ip^2 \sigma_X \otimes \sigma_X -i \sigma^1_Y \Big)\frac{\partial}{ \partial p^1}
    \end{split}
  \right\}
\end{equation*}
The sectional curvature is then \[\displaystyle \kappa = \frac{8}{9} \frac{(p^1)^2 + (p^2)^2}{(p^1)^2 + (p^2)^2+1}.\] The non-constant curvature with such a simple example is quite curious, but not too surprising. Given the strange nature of entanglement and, consequently, non-commutativity of an arbitrary vector space, one might expect a strange curvature.  Interestingly, though, all curvature is strictly positive.  

For completeness, taking the perspective of the dynamic Lie algebra, the root elements in the Lie derivative, derived through multiple iterations of the commutator, are $\Big\{i  Y \otimes I, i  I \otimes Y,  i  X \otimes X, i X \otimes Y, i  Z \otimes X, i  Z \otimes Z \Big\}$. Collecting the coefficients for each root is quite unwieldy, and so, not given. For the sectional curvatures, out of the possible $15$ directions, $5$ were zero. Therefore, the curvature is non-constant and not necessarily positive. Hence, just like the coordinate orthonormal basis, the manifold is not Einstein, but locally dependent. 

The dichotomy between the coordinate vector field and the dynamic Lie algebra requires further investigation. The following section dives deeper into this interplay.       

\section{\label{sec:cotangent-vol}Volume and Harmonic Maps}

The concepts of volume, energy density, and harmonic functions are well-established; see Jost Chapter 9 \cite{jost2008riemannian}. However, the subtleties with the root elements of our vector space need to be addressed to adjust the formal definitions. Furthermore, from our previous work, we have that the quantum feature map is an \textit{immersion}. Therefore, we may leverage the pullback measure to establish many of the formulas.

Recall that for a Hamiltonian quantum feature map $U:M \to \SU(2^N)$ as in Equation \eqref{eq:faeture-map} and Assumption \ref{ass:general-map}, the differential operator that acts on $M$ has closed form 
\begin{equation}\label{eq:generator-vector-field}
\displaystyle \sum_{i=1}^{m} \left(  \sum_{k=1}^{n} \frac{\partial f_{k}}{\partial x^i} e^{L(\cdot)}\left( \frac{\I - e^{-\mathrm{ad}_{L(\cdot)}}}{\mathrm{ad}_{L(\cdot)}} \right) L_k \right) \frac{\partial }{\partial x^i},
\end{equation}
where $\displaystyle \frac{\partial }{\partial x^i}$ is taken to be the differential operator, allowing for incorporation of any smooth path on $M$. For generality, Equation \ref{eq:generator-vector-field} will be used for the analysis. Now, observe for the subset of root elements that generate the dynamic Lie algebra $\mathfrak{su}_{L(\cdot)}$, $\{E_1, \ldots, E_{n'}\}$, we may combine all coordinates and write the equation above in the form
 \begin{equation}\label{eq:vector-field}
 \begin{split}
     &  \sum_{i=1}^{m}\left( \sum_{k=1}^{n'}C^i_k e^{L(\cdot)} E_k \right)  \frac{\partial }{\partial x^i} 
      =  \sum_{i=1}^{m} \sum_{k=1}^{n'} e^{L(\cdot)} E_k \left( C^i_k\frac{\partial }{\partial x^i} \right) \\ 
      & =  \sum_{k=1}^{n'}  \left( \sum_{i=1}^{m} C^i_k\frac{\partial }{\partial x^i} \right) e^{L(\cdot)} E_k 
      =  \sum_{k=1}^{n'}  \left( \sum_{i= 1}^{m} C^{i_k} \frac{\partial }{\partial x^{i} } \right) e^{L(\cdot)} E_k.
 \end{split}  
 \end{equation}
Note that within the sum of the partials the functions $C^{i}_{k}$ are smooth and scalar-valued with domain $M$, and with the potential for some $C^{i}_k \equiv 0$. 

\begin{remark}
    From Equation \eqref{eq:vector-field}, one may see that $\Big\{  e^{L(\cdot)}E_1 , \ldots,  e^{L(\cdot)}E_{n'} \Big\}$ is a \define{frame}, and hence, $\Big\{  
        g\big( e^{L(\cdot)}E_1, \cdot \big) , \ldots, g\big( e^{L(\cdot)} E_{n'}, \cdot \big) \Big\} $ is a \define{coframe}. 
\end{remark}

Quite interestingly, Equation \eqref{eq:vector-field} displays how the change of direction for an arbitrary $p\in M$ dictates each root's local change of direction. This has an effect on analytic forms of volume and harmonic maps, which is shown in the following subsection.

\subsection{Volume and Energy}

We proceed to define the \define{volume} of the space $U(M)$. Recall that for a Riemannian manifold and metric $(H,\alpha)$ of degree $h$,  for $\sqrt{\alpha}:= \sqrt{|\mbox{det}\big( \alpha(\partial_i,\partial_j) \big)|}$ and the basis of cotangent bundle $\big\{dx^i \big\}_{i=1,\ldots,h}$, the standard formula for the volume is given as
\begin{equation}\label{eq:volume-general}
    \mbox{Vol}(H) = \int_{H} \sqrt{\alpha(x)} dx^1 \wedge dx^2 \wedge \cdots \wedge dx^h.
\end{equation}
As noted in Petersen \cite{petersen2006riemannian}, this is for notational convenience and should not be understood as the exterior derivative of an $(n-1)$-form. 

Taking this note and coupling it with the pullback metric from the map $U$, one may observe that the formula for the volume is the form  
\begin{equation}\label{eq:volume-su}
   \displaystyle \mbox{Vol}\big( U(M) \big) = \int_{M} \prod_{k=1}^{n'}\bigg| \sum_{i=1}^{m_k}C^{i}_k(x) \bigg| dx^{1} \wedge \ldots \wedge dx^m.
\end{equation}
Specifically, recall that for a map between to manifolds $\Phi:(Q,q) \to (S,s)$ then the formula for the pullback metric is $\displaystyle \Phi^*s = s_{\alpha,\beta} \sum_{i,j}\frac{\partial \Phi^{\alpha}}{\partial x^i} \frac{\partial \Phi^{\beta}}{\partial x^j}$; see Jost \cite{jost2008riemannian} for subtleties, therein. From this formula, we connect the subscribes of $\alpha$ and $\beta$ to the elements in the Equation \eqref{eq:vector-field}, which then yields the matrix  
\begin{equation*}
\left( g \left( \left( \sum_{i=1}^{m_j} C^i_j(x) \right)  e^{L(x)}E_j , \left( \sum_{i=1}^{m_k} C^i_k(x) \right)  e^{L(x)} E_k  \right) \right)_{j,k},
\end{equation*}
and simplifies to a diagonal matrix with respective squared sums.  

Now that we have derived a closed form for the volume, we may move on to find closed forms for the energy density and the energy map or energy functional. For the energy density, Chapter 9 in Jost \cite{jost2008riemannian} gives a clearer and more general form, which we will apply to the focused encoding scheme. 

For two Riemannian manifolds and a $C^1$ function, $W:(Q,q) \to (S,s)$, the \define{energy density} is defined as
\begin{equation}
    e(W) := \frac{1}{2} \big\langle dW, dW \big\rangle_{T^*Q \otimes W^*TS} \ .
\end{equation}
With this definition, one may show that 
\begin{equation}\label{eq:energy-density}
    e(U) = \frac{1}{2} \sum_{\alpha, \beta = 1}^{m} l^{\alpha \beta} g\left( \frac{\partial U}{\partial x^{\alpha}} , \frac{\partial U}{\partial x^{\beta}} \right),
\end{equation}
where the superscript indices on the metric $l$ indicate the (matrix) inverse, and $\displaystyle \frac{\partial U}{\partial x^{\alpha}}$ are the respective terms in Equation \eqref{eq:vector-field}.

Following the logic in Section 9.1 in Jost \cite{jost2008riemannian} and incorporating the roots elements $\big\{ E_k \big\}_{k=1,\ldots,n'}$ within the arguments will yield the simplification in Equation \eqref{eq:energy-density}.  

From the energy density and volume, the definition of the \define{energy map} of $M$ to $\SU(2^N)$ is
\begin{equation} \label{eq:energy-map}
\begin{split}
       E(U) := \int_{M} e(U)(x) dVol(M).
\end{split}
\end{equation}
Combining Equation \eqref{eq:volume-su} and Equation \eqref{eq:energy-density} yields a closed formula for computation, albeit, some work is required to simplify the energy density function.

\subsection{\label{subsec:harmonic-maps}Harmonic Maps}
The energy map defined in Equation \eqref{eq:energy-map} contains a lot of information connecting the two manifolds of $M$ and $U(M)$. Specifically, the values of $E$ yield information about how naturally a manifold maps to another with large values indicating significant distortion, and small values indicating a similar structure. Of course, the existence of critical points are of extreme importance. 

This subsection is essentially takes the details in Chapter 9 in Jost \cite{jost2008riemannian} and gives an analytic form of a harmonic map for a Hamiltonian quantum feature map. While this mainly consists of filling in rigorous details, the purpose is to mitigate any confusion with the general analytic form for a harmonic map to ensure ease of computational or theoretical use in applications.

Maps that are critical points are denoted as \define{harmonic maps} and, for our focus on Hamiltonian feature maps, requires adjustment from the standard definition. For harmonic maps, a one-parameter evolution, say $t$, is applied to the energy map and the maps where the derivative at $t=0$ is $0$ are called harmonic maps; for deeper detail, please see Jost \cite{jost2008riemannian} and Robbin and Salamon \cite{robbin2022introduction}. Following the logic of the proof in Lemma 9.1.1 in Jost \cite{jost2008riemannian} and denoting $\displaystyle \left( \frac{\partial U}{ \partial x^{\alpha}} \right)^i$ as the function-coefficient of the $i^{th}$ root $E_i$, then we have the analytic form  
\begin{equation}\label{eq:euler-lagrange}
   \sum_{\alpha, \beta = 1}^{m}  \left( 
   \begin{split}
        & \frac{ 1 }{ \sqrt{l} } \sum_{i = 1}^{n'}   \frac{ \partial }{ \partial x^{\alpha} } \left( \sqrt{l} l^{\alpha \beta} \left( \frac{ \partial U }{ \partial x^{\beta} } \right)^i  \right) \\
        + & l^{\alpha \beta} \sum_{j = 1}^{n'}  \left( \frac{ \partial U}{ \partial x^{\alpha} } \right)^j  \left( \frac{ \partial U}{ \partial x^{\beta} }\right)^j   
   \end{split}             
   \right) = 0.
\end{equation}
Equation \eqref{eq:euler-lagrange} is called the \define{Euler-Lagrange} equation.

 \begin{remark}\label{re:euler-lagrange-deeper}
     Observe here that one would have expected to see the \define{Christoffel symbol} $\Gamma^{i}_{jk}$ in Equation \eqref{eq:euler-lagrange}. However, expanding the term metric term in the energy density simplifies to $\displaystyle g\left( \frac{\partial F}{\partial x^{\alpha}} , \frac{\partial F}{\partial x^{\beta}} \right) =  \sum_{j = 1}^{n'}  \left( \frac{ \partial F}{ \partial x^{\alpha} } \right)^j  \left( \frac{ \partial F}{ \partial x^{\beta} }\right)^j $. Hence, many terms in the standard form for the Euler-Lagrange equation naturally disappear. 
 \end{remark}

To be more formal and tangible, we will give a specific definition for a critical point. For this definition, we need the concept of an \define{exponential map}. Exponential maps would have been introduced earlier, using the flow in the standard literature, but it would have been awkwardly placed. The following definition, while standard, is taken from Jost \cite{jost2008riemannian}.

\begin{definition}\label{def:exp-map}
    Let $M$ be a Riemannian manifold, $p \in M$, $c_v$ a geodesic such that $c_v(1) = p$ and $\dot{c}_v(0)=v$, $V_p := \left\{ v \in T_p M : c_v \mbox{ is defined on } [0,1] \right\}$, and $\exp_p : V_p \to M$ where $v \mapsto c_v(1)$. Then $\exp_p$ is called the exponential map. 
\end{definition}

Notice that an exponential map is generated by a vector field. Thus, for $\psi$ a vector field along $U$---ergo, $\psi$ is a vector field in $U^*T\SU(2^N)$---then $\psi$ induces a variation of $U$ by 
\begin{equation}\label{eq:variation-exp}
    U_t(x):=\exp_{U(x)}\Big(t \psi(x) \Big).
\end{equation}

We are now ready to state the formal definition of a critical point.
\begin{definition}\label{def:critical-point}
    A smooth function $U$ is a critical point of the energy map $E$ if, whenever $\psi$ is a compactly supported bounded vector field of $U^*T\SU(2^N)$, we have 
    $$
    \frac{d}{dt}E(U_t) \Big|_{t=0} = 0.
    $$
    In this context, compactly supported means that $\psi$ is in the Sobolev space $H^{1,2}\big(M,\SU(2^N) \big)$, ergo, $\displaystyle \int_{M}\langle d\psi, d\psi \rangle dVol(M) < \infty$. 
\end{definition}

This leads us to an important lemma linking critical points to exponential variation, and can be found in Jost \cite{jost2006compact}. 
\begin{lemma}
    We have $F \in H_{ \mbox{loc} }^{1,2}$ is a critical point iff 
    $$
    \displaystyle  \int_{M} g\big( dF , d\psi \big) dVol(M) = -\int_{M} g\big( \Tr \nabla dF , \psi \big) dVol(M)=0,
    $$
    where $\psi$ is as in Definition \ref{def:critical-point}.
\end{lemma}
The term $\tau(F):= \Tr \nabla dF$ is called the \define{tension field} of $F$, and one may see that the Euler-Lagrange equations for $E$ is equal to $\tau$ up to a sign. Hence $\tau(F) = 0$ if and only if $F$ is harmonic. Moreover, one may use Equation \eqref{eq:euler-lagrange} to determine if a given function is a critical point. There is a closed form for $d\psi$ in Jost, but is a bit more subtle and not discussed.   

Of course, one may extend the function $F$ to be in the class of Hamiltonian feature maps, $F:M \to \SU(2^{\eta})$, and work directly with encoding schemes and work from the ground-up. This is expected to be a quite difficult task as one needs to consider the value and effect of $\eta$, the smooth functions $f_k$, the elements $L_k$, and the final vector field given by Equation \eqref{eq:vector-field}. This paramount classification is left for another paper.

\section{\label{sec:discussion}Discussion}
\subsubsection{Summary}
This manuscript rigorously established a Riemannian manifold structure on the range of a class of quantum feature maps, taken to be a subset of the Lie group $\SU(2^N)$. Since this subset is only a Lie subgroup, or even just a subgroup, in extremely rare scenarios, the establishment required a ground-up approach, working from base-level definitions then showing assumption of lemmas and propositions hold to verify analogous results from the Riemannian structure on matrix Lie groups. 

One of the more important results shows that geodesic on the embedded manifold is also the geodesic on the range of the quantum feature map. Thus, considerably simplifying the computation of geodesic distance. This also includes manifolds without a known Riemannian manifold structure; think about a point cloud and taking the assumption the data points are representative of a manifold. In other words, it takes less resources to determine a geodesic on an embedded manifold than on a Lie group of matrices.   

From the metric, we established a means to calculate curvature, and gave an example to illustrate subtleties of this calculation. The most import piece of the calculation was the corroboration of an explicit Levi-Civita connection. 

Lastly, we established the closed forms for the volume, energy map, and formula for a harmonic function. These derivation of the yielded significant insight into the interaction of the base manifold and the effects of choosing the scalar-valued functions and skew-Hermitian matrices. While these results are a natural progression in developing a rigorous framework for a Riemannian manifold, the closed analytic form to identify harmonic functions will be essential in deriving information loss.

\subsubsection{Future Directions}
The characterization of information loss of different classes of manifolds and respective quantum feature maps is an essential to further obtain rigorous insight into selecting encoding schemes.  As a starting point, the information obtained from the energy map and the critical points, or harmonic functions, should indicate how the research will evolve. Generally, the Riemannian geometry structure derived in this paper can be leveraged to derive a quantum feature map that retains the geometric structure of of the base manifold. Or mathematically show such a map does not exist with a given manifold. Of course, one may foresee that further mathematical exploration will be necessary, which includes information from the characteristic classes of Euler and Pontryagin.

Consequently, this research path should also yield deeper insight in the sequential layers on the quantum circuit. In fact, it is posited that there is a connection between the geometric structure of an encoding scheme and the performance of a quantum-based statistical model. Particularly, the deformation of the geometric structure of an embedded manifold to $\SU(2^N)$ will indicate performance of a particular task. For instance, as was shown, a flat structure indicates no quantum advantage. Moreover, IQP has been shown to generate an exponential concentration of states, and the curvature of a simple example was shown to be non-constant. These observations are fairly superficial, and further research of this Riemannian manifold structure should yield deeper insight to circuit derivation.

\begin{acknowledgments}
The author would like to thank Arthur Parzygnat and Dan Ferrante for many helpful discussions and guidance. 
\end{acknowledgments}

\bibliographystyle{unsrt}
\bibliography{refs}
\end{document}